# Glucose Sensing Using Pristine and Co-doped Hematite Fiber-Optic sensors: Experimental and DFT Analysis

Namrata Pattanayak[a,b,*], Preeti Das[c], Mihir Ranjan Sahoo[d], Padmalochan Panda[e], Monalisa Pradhan[f], Kalpataru Pradhan[g], Reshma Nayak[a], Sumanta Kumar Patnaik[a], Sukanta Kumar Tripathy[c]

[a] Department of Physics, NIST University, Berhampur 761008, India.

[b] Institute of Materials for Electronics and Magnetism (IMEM), National Research Council (CNR), Parco Area delle Scienze 37/A, Parma 43124, Italy.

[c] Centre Of Excellence in Nanoscience and Technology for the Development of Sensor, P.G. Department of Physics, Berhampur University, Bhanja Bihar, Odisha, 760007, India

[d] Graz University of Technology, Graz, 8010, Austria

[e] Center for Biomolecular Nanotechnologies, Istituto Italiano di Tecnologia, 73010 Arnesano, Lecce, Italy

[f] Department of Physics, School of Applied Sciences, KIIT Deemed to be University, Bhubaneswar 751024, India

[g] Theory Division, Saha Institute of Nuclear Physics, A CI of Homi Bhabha National Institute, Kolkata 700064, India

[*] namratapattanayak@gmail.com

## Abstract

Glucose monitoring plays a critical role in managing diabetes, one of the most prevalent diseases globally. The development of fast-responsive, cost-effective, and biocompatible glucose sensors is essential for improving patient care. In this study, a comparative analysis is conducted between pristine and Co-doped hematite samples, synthesized via the hydrothermal method, to evaluate their structural, morphological, and optical properties. The glucose sensing performance of both samples is assessed using a fiber-optic evanescent wave (FOEW) setup. While the sensitivity remains comparable for both pristine and Co-doped hematite, a reduction in the Limit of Detection (LoD) is observed in the Co-doped sample, suggesting enhanced interactions with glucose molecules at the surface. To gain further insights into the glucose adsorption mechanisms, Density Functional Theory (DFT) calculations are performed, revealing key details regarding charge transfer, electronic delocalization, and glucose binding on the hematite surfaces. These findings highlight the potential of Co-doped hematite for advanced glucose sensing applications, offering a valuable synergy between experimental and theoretical approaches for further exploration in biosensing technologies.

## 1. Introduction

Glucose, as a vital component of cellular energy, plays a pivotal role in sustaining human life and health. Maintaining stable blood glucose levels is essential for overall well-being, as any imbalance can lead to serious health conditions such as diabetes. According to the World Health Organization (WHO), diabetes is one of the most frequently diagnosed diseases globally, with projections suggesting that by 2030, the number of individuals affected could reach 578 million [1, 2]. As a result, there has been a growing interest in developing fast-responsive and convenient glucose biosensors to monitor blood glucose levels effectively and help manage the increasing burden of diabetes, with applications extending beyond medical science to include the food industry as well [3, 4, 5, 6, 7, 8].

To meet these demands, various glucose detection methods have been developed, including colorimetry, chromatography, fluorescence, Raman spectroscopy, and electrochemical techniques [9, 10, 11, 12, 13]. Alongside these techniques, various oxide semiconductors such as $BiVO_4$, $TiO_2$, $Co_3O_4$, $WO_3$, ZnO, and α-$Fe_2O_3$ (hematite) have been explored with the aim of improving biosensor performance namely, the sensitivity, selectivity, and response time of glucose biosensors [8, 14, 15, 16, 17]. Among these materials, hematite stands out due to its earth abundance, ease of fabrication in various morphologies, cost-effectiveness, biocompatibility, and biodegradability, making it a promising candidate for in-vivo biosensing and bio-detection applications [16, 17, 18, 19, 20, 21, 22, 23]. Hematite's tunable optical, electronic, and catalytic properties make it particularly appealing for the development of various sensing devices, including glucose biosensors [19, 23, 24, 25]. In particular, glucose sensing using hematite has been predominantly explored through photoelectrochemical (PEC) methods due to hematite's favorable photoactive properties, but its potential for other sensing techniques remains underexplored [16, 17, 18, 19, 20, 21, 22, 23].

Recently, fiber-optic evanescent wave (FOEW)-based methods have emerged as promising tools for developing non-invasive and non-enzymatic glucose detection systems [25, 26, 27, 28]. These methods are highly attractive for their potential to offer low-cost, reduced-complexity, and improved patient compliance in biosensing devices. More-over, optical fiber sensors can be designed with geometrical versatility, allowing configurations like straight, bent, or flattened fibers to optimize sensitivity [26, 27, 28, 29, 30]. In addition, FOEW sensors coated with nanoparticles offer enhanced sensitivity, biocompatibility, and ease of integration into optical systems, making them an appealing alternative to traditional glucose sensors [25, 26, 27, 28, 29, 30, 31]. In these sensors, the evanescent field generated by light traveling through the optical fiber interacts with nanoparticles on the fiber's surface, which selectively bind glucose molecules from the surrounding medium. This interaction causes changes in the light's intensity or wavelength, which are detected and quantified to determine glucose concentration.

Although FOEW-based glucose sensors coated with oxide semiconductors such as ZnO and $Co_3O_4$ have been extensively studied [31, 32, 33, 34, 35], the application of hematite-coated FOEW biosensors for glucose detection remains relatively unexplored. Cobalt doping, in particular, has shown potential in tuning the band gap and enhancing the charge transfer properties of hematite, which could improve its glucose sensing performance [36, 37, 38, 39]. In this study, we investigate both pristine and Co-doped hematite for their potential application in glucose sensing. The hematite samples are synthesized using the hydrothermal method and characterized through Rietveld refinement of X-ray diffraction (XRD), scanning electron microscopy (SEM), and UV-Visible spectroscopy to assess their structural, morphological, and optical properties. The glucose sensing performance of both the pristine and Co-doped hematite samples is evaluated using the FOEW detection method in a homemade setup. To the best of our knowledge, such an attempt of glucose detection using the hematite-coated FOEW method has not been reported yet and is important from an experimental perspective.

In addition to the experimental investigation, our work also includes a comprehensive Density Functional Theory (DFT) analysis to provide deeper insights into the electronic structure and interaction mechanisms between glucose molecules and hematite surfaces. The DFT calculations revealed key aspects of glucose adsorption, electronic delocalization, and charge transfer, highlighting the potential of Co-doped hematite as an advanced material for glucose sensing applications.

## 2. Experimental

### 2.1. Synthesis of pristine hematite and Co-doped hematite pseudocubes

Chemicals sourced from Merck, graded analytically, are employed in the hydrothermal synthesis of pristine and Co (2%)-hematite pseudocubes. For the typical synthesis of Co (2%)-hematite, the precursors $FeCl_3.6H_2O$ and $CoCl_2.6H_2O$ in appropriate molar ratios, are first dissolved in 16 ml of deionized distilled water (DDW). A separate solution is prepared by dissolving 0.7 g of NaOH pellets in 16 ml DDW. Following thorough mixing under sonication, the combined solutions are transferred into a Teflon-lined stainless-steel autoclave. The reaction is carried out for 12 hours at a temperature of 160 °C. The resulting precipitate, deposited at the bottom of the Teflon cup, is collected through centrifugation, extensively washed using DDW and ethanol, and finally left to air dry at room temperature for several hours before characterization. The synthesis protocol for pristine hematite follows a similar procedure, but without the incorporation of a cobalt source.

### 2.2. Characterization details

The phase purity and crystallinity of the samples have been investigated using a powder X-ray diffractometer, Bruker D8 ADVANCE. Cu Kα radiation (λ=1.54178 Å) with a tube operating voltage of 40 kV and current of 30 mA is used to record the diffraction patterns. The diffraction patterns are further examined using a computer-based Rietveld refinement technique. Morphological analysis of the samples has been carried out using a Zeiss$^{TM}$ Ultra plus field-emission scanning electron microscope (FESEM). An energy dispersive analysis of X-ray (EDAX) detector equipped with FESEM is utilized for the elemental/chemical composition analysis. The absorption spectroscopy measurements have been carried out using UV-Visible (Agilent Cary-3500) spectrophotometer.

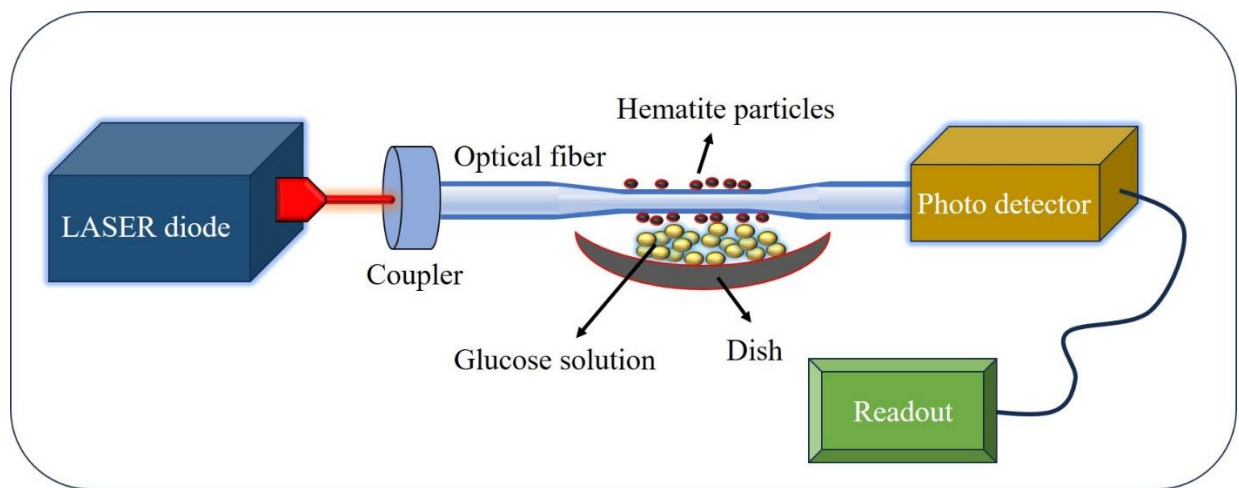

*Figure 1: Schematic of the FOEW sensor setup for glucose sensing.*

*2.3. Preparation of fiber-optic sensing probe and measurement*

Figure 1 depicts the schematic of the FOEW sensor setup used in this study. The setup consists of several components: a laser diode emitting light with a wavelength of 633 nm, a coupler, a poly(methylmethacrylate) optical fiber measuring 250 μm in diameter and 30 cm in length, and a photodetector.

The glucose sensing region is developed by flattening a 2 cm segment from the center of the optical fiber, denoted as $F_{flatten}$, for experimentation. Flattening this segment increases the fraction of the evanescent field, enhancing the fiber's sensitivity to changes in the surrounding glucose concentration [40, 41, 42]. Additionally, the glucose sensing region is sensitized by coating it with synthesized hematite particles. For coating, two separate solutions are prepared by dissolving 5 mg of each sample in 2 ml of Milli-Q water. The flattened fibers, after being cleaned several times with Milli-Q water, are dip-coated using gradual heat and cool mechanism with the prepared solution and dried at room temperature inside the desiccator. A set of fibers is prepared for the sensing application using a similar method. The final sensing probes, after the flattened fibers coated with pristine and Co-doped hematite, are named $F_{pristine-hematite}$ and $F_{Co-doped-hematite}$ respectively.

A typical healthy individual's blood glucose levels range from 80 to 120 mg/dL (4.4 to 6.6 mM). In consideration of this, various concentrations of high purity (>99%) glucose solution (ranging from 0 mM to 11 mM) are separately prepared by the serial dilution method. The sensing region of the prepared probes is then dipped into a Petri dish containing the glucose solution, as shown in Figure 1. The laser light is directed through a suitable coupler to enter the fibers, and the light reaching the other end of the fibers is directed onto a photodetector. This photodetector is connected to a measuring unit, which records the output current in microamperes (μA).

## 3. Results and Discussions

*3.1. Structural and morphological analysis*

In Figure 2(a, b), FESEM micrographs of both pristine and doped hematite samples are depicted. These micro-graphs reveal that the individual grains of both the samples are pseudocubic in shape and possess a porous/rough surface morphology. The side length of the pseudocubes measures approximately 1 μm for both pristine and doped hematite. Figure 2(c) and (d) display X-ray diffraction (XRD) patterns of both pristine and doped hematite samples, alongside Rietveld fitting using the FULLPROF suite. The fitting process involves step by step refinement of diffraction profile parameters to align the calculated profile ($I_{calc}$) with the experimental profile ($I_{obs}$). Fitting progress is assessed by comparing numerical R values, with a $\chi^2$ ($=(R_{wp}/R_{exp})^2$) value approaching 1 indicating a good fit. It is crucial to emphasize that, beyond numerical R values, a graphical visualization of the observed and calculated patterns plays a significant role in evaluating the Rietveld fit's quality [43, 44, 45]. From Figure 2(c) and (d) we observe a reasonably good agreement between the observed and calculated patterns, affirming that all the samples stabilized in the hexagonal R3-c symmetry of hematite without any impurities or secondary phases. The structural and goodness-of-fit (GOF) parameters of the pristine and doped hematite samples are summarized in Table 1. From Table 1, the chemical composition of the doped hematite sample is found to be $Fe_{1.96}Co_{0.04}O_3$ which has further been corroborated by EDAX analysis.

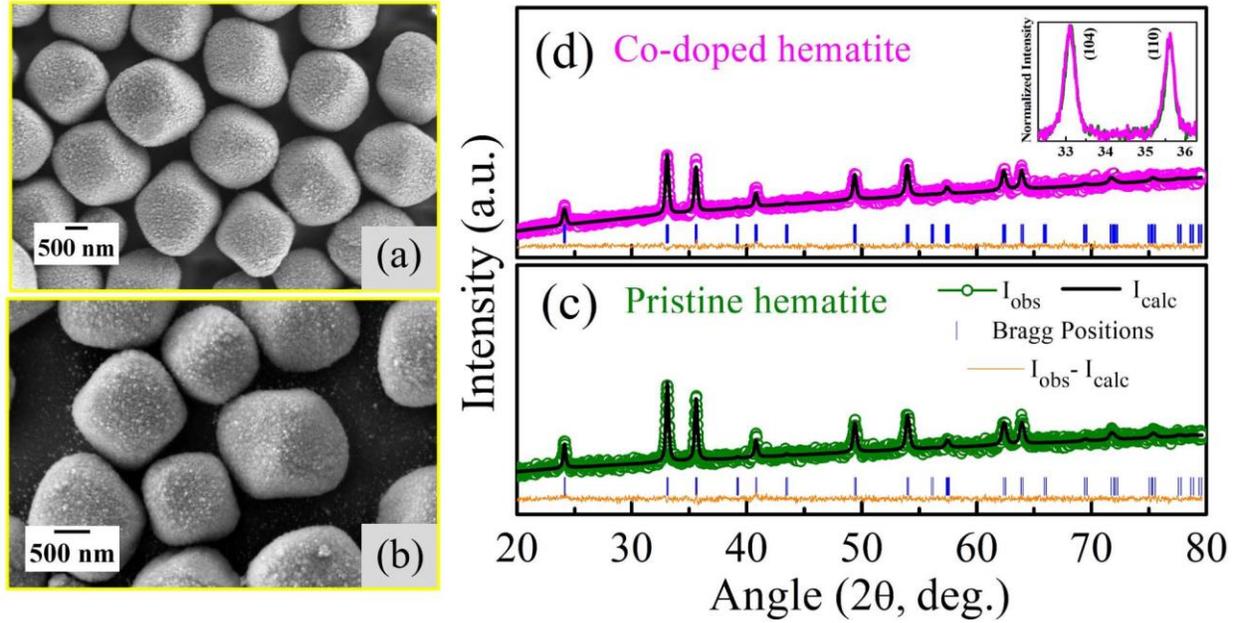

*Figure 2: SEM micrograph of (a) pristine hematite and (b) Co-doped hematite pseudocubes. X-ray diffraction data with Rietveld fit carried out using FULLPROF suite for (c) pristine hematite and (d) Co-doped hematite.*

| Sample | Lattice constants (Å) | | Unit cell volume (Å)$^3$ | Atom | Wickoff | x/a | y/b | z/c | occupancy | GOF parameters | | |
|---|---|---|---|---|---|---|---|---|---|---|---|---|
| | a=b | c | | | | | | | | $R_{wp}$ | $R_{exp}$ | $\chi^2$ |
| Pristine hematite | 5.0422 (0.0006) | 13.7865 (0.0021) | 303.547 (0.068) | Fe | 12c | 0 | 0 | 0.35530 | 0.33333 | 2.20 | 2.15 | 1.05 |
| | | | | O | 18e | 0.30590 | 0 | 0.25 | 0.50000 | | | |
| Co-doped hematite | 5.0436 (0.0003) | 13.7961 (0.0014) | 303.925 (0.041) | Fe | 12c | 0 | 0 | 0.35517 | 0.32683 | 2.47 | 2.33 | 1.12 |
| | | | | O | 18e | 0.30188 | 0 | 0.25 | 0.50010 | | | |

*Table 1: Structural parameters of pristine and Al doped hematite samples as determined from the Rietveld analysis of the room temperature x-ray diffraction data. (The atomic site occupancy is defined here as the ratio of site multiplicity to the general multiplicity of the $R\bar{3}c$ symmetry of hematite)*

In general, introducing a dopant with an ionic radius different from the host lattice can lead to shifts in XRD peaks, alterations in lattice parameters, and changes in unit cell volume [39, 44, 46, 47]. Nonetheless, the extent of strain generated in a doped sample depends on dopant concentration. Additionally, for a given dopant concentration, the degree of strain induced in the lattice depends on the dopant size [39, 44]. From the inset of Figure 2(d), we observe no significant shift in XRD peaks of the doped hematite sample compared to the pristine one. This consistency is also reflected in the lattice parameter and unit cell volume of the doped hematite, showing no substantial deviation from pristine hematite, as demonstrated in Table 1. To further analyse the effect of Co-doping on the hematite crystal, we estimated the crystallite size and micro-strain of both pristine and doped hematite samples using the Williamson-Hall (W–H) method depicted in Figure 3 [44, 48, 49]. We deconvoluted the contribution from instrumental broadening using peak broadening of standard Silicon material. The calculated crystallite sizes for pristine and doped hematite are 46 nm and 44 nm, respectively. Similarly, we found that the strain is tensile in nature, with calculated values of $1.8 \times 10^{-3}$ and $1.9 \times 10^{-3}$ for pristine and doped hematite samples, respectively. The present findings suggest that the partial substitution of high-spin $Fe^{3+}$ ions (ionic radius 64.5 pm) with low-spin $Co^{2+}$ ions (ionic radius 65 pm) in the hematite lattice, thereby leading to minimal lattice strain [39, 50].

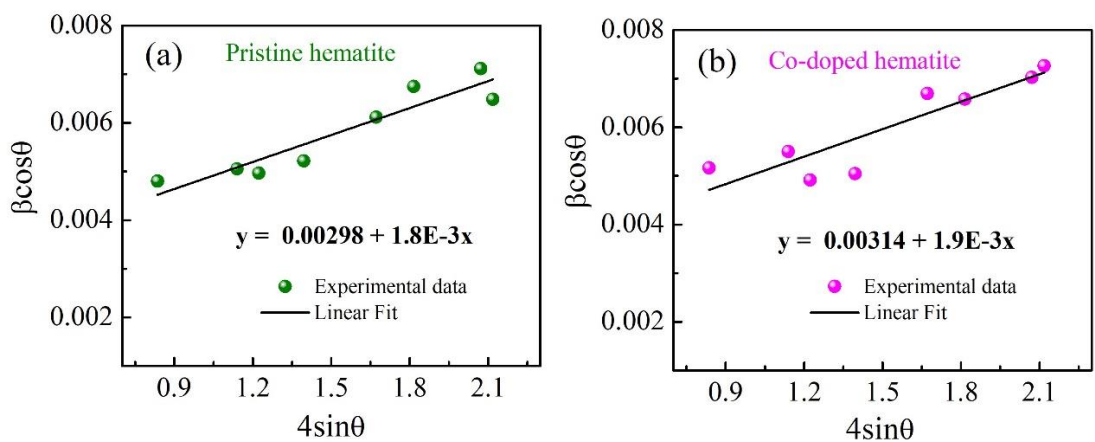

*Figure 3: Williamson-Hall plots for (a) pristine hematite and (b) Co-doped hematite.*

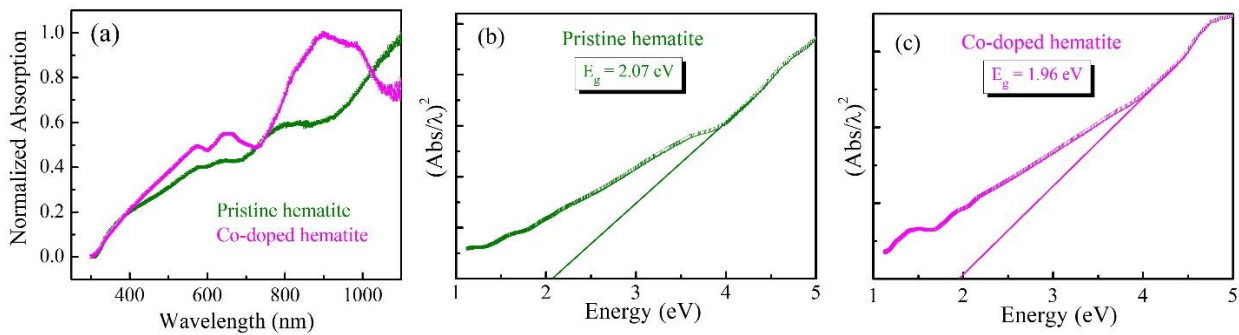

*Figure 4: (a) Normalized UV-Visible absorption spectra of pristine and Co-doped hematite, showing distinct ligand field transitions (LFT) and shifts in absorption features. Tauc plots showing estimated optical bandgaps for (b) pristine hematite and (c) Co-doped hematite.*

### 3.2. UV-Visible analysis

The optical properties of both pristine and Co-doped hematite are investigated using UV-Visible spectroscopy in the range of 300–1100 nm. As shown in Figure 4(a), the spectra reveal distinct absorption features characteristic of hematite [39, 51, 52, 53, 54]. In particular, the ligand field transitions (LFT) dominate in the 700–900 nm region for pristine hematite and in the 750–1000 nm region for the Co-doped sample, reflecting the influence of microcube morphology on light absorption [53, 54]. Additionally, two smaller humps are observed in the 550–700 nm range for both samples, which can be attributed to $Fe^{3+}$ ligand field transitions and pair excitations involving magnetically coupled ions [39, 51, 52, 53, 54]. A red shift and enhanced intensity in the Co-doped sample in these regions suggest modifications in the local electronic environment due to cobalt doping.

The optical bandgaps of the prepared hematite and Co-doped hematite samples are determined using a modified form of the Tauc relationship, where $(Abs/\lambda)^2 = B(h\nu - E_g)$, with B being a constant related to the shape and size of the nanostructures [41, 55]. A plot of $(Abs/\lambda)^2$ versus energy $(h\nu)$ is generated, and the bandgap is obtained by extrapolating the linear portion of the curve to the x-axis, where $(Abs/\lambda)^2 = 0$, corresponding to the bandgap energy. As depicted in Figures 4(b) and 4(c), the optical bandgaps are found to be approximately 2.07 eV for pristine hematite and 1.96 eV for Co-doped hematite, indicating a narrowing of the bandgap. This narrowing in the bandgap can be attributed to the introduction of additional electronic states by cobalt, which facilitates light absorption in the visible and near-infrared regions.

### 3.3. Glucose sensing performance and specificity analysis

For the glucose detection experiment, we first measured the detector output current for the $F_{flatten}$ configuration in the absence of any analyte, which is found to be 39.2 μA. We then tested whether there are any differences in the detector output when the flattened portion of the fiber is dipped into different concentrations of glucose solution. No appreciable change is observed in the detector output for the $F_{flatten}$ configuration. However, a slight but noticeable variation in the detector output is observed with a change in glucose concentration for both the $F_{pristine-hematite}$ and $F_{Co-doped-hematite}$ probes, each with response times of 60 seconds. As shown in Figure 5(a) and 5(b), the detector output current varies linearly with glucose concentrations ranging from 0 to 11 mM for both probes. Notably, the upper and lower current ranges for both probes remain similar. The experimental data points are fitted linearly, and the slope of the linear fit equation indicated the sensitivity of the probes.

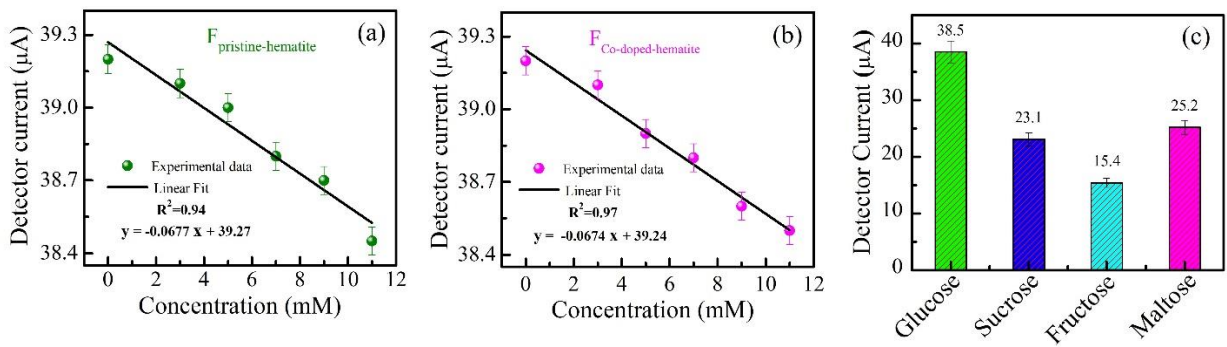

*Figure 5: (a) Detector output current variation with glucose concentration for the Fpristine-hematite probe configuration. (b) Corresponding output current response for the FCo-doped-hematite probe. (c) Comparative detector output current response of the FCo-doped-hematite probe at 11 mM concentration for various interfering species.*

The limit of detection (LoD) of the probes is evaluated using the formula:

$$LoD = \frac{3.3 \times S.D \; of \; intercept}{slope}$$

where S.D is the standard deviation of the intercept [31]. After repeated measurements, we observed that the sensitivity values for pristine and Co-doped hematite samples remained comparable within the margin of error, with the pristine sample showing -0.0677±0.0074 and the Co-doped sample -0.0674 ±0.0048. However, the LoD is notably different for both probes, with values calculated as 6.12 mM for $F_{pristine-hematite}$ and 3.99 mM for the $F_{Co-doped-hematite}$ configurations.

The absence of a noticeable change in sensitivity between the two probes suggests that the interaction of the evanescent wave across the coating material and coating material-glucose analyte interfaces is largely the same for both probes. While the 2% Co-doping induces a shift in the absorption edge and narrows the bandgap, this may not significantly modulate the refractive index difference between the Co-doped and pristine hematite when interacting with glucose [31, 29, 25, 40]. It is also important to note that particles with a higher surface-to-volume (S/V) ratio generally allow more glucose molecules to be adsorbed onto the surface, thereby improving sensor performance [31, 29, 25]. However, from the SEM micrographs, the amount of glucose adsorbed onto both the pristine and Co-doped hematite surfaces appears comparable, likely due to the similar overall surface areas accessible for glucose adsorption.

While the sensitivity values remain comparable, the LoD is notably different for the two probes. The lower LoD observed in the Co-doped hematite sample can be attributed to the enhanced interaction strength between the Co-doped hematite surface and glucose molecules, a result further substantiated by the theoretical analysis using DFT calculations, as discussed in the subsequent section. Although this interaction does not lead to a significant change in sensitivity, it is sufficient to improve the sensor's ability to detect lower glucose concentrations. The refractive index at the interface between the fiber surface and the analyte likely plays a more significant role in determining sensitivity than surface interaction alone [31, 29, 25].

Another important characteristic of a reliable sensor is its specificity towards the targeted analyte. To verify the specificity of the hematite-based glucose sensor, we extended our study to test its response in the presence of common interfering species, such as fructose, maltose, and sucrose, which are often found in glucose-containing solutions. Since the $F_{Co-doped-hematite}$ probe showed improved Limit of Detection (LoD), the specificity experiments are conducted solely on this probe. The same experimental procedure was followed to ensure that these interfering species did not affect glucose detection. The results demonstrated that the presence of these interfering species have no significant effect on the output current, confirming the high specificity of the $F_{Co-pristine-hematite}$ probe towards glucose detection, as shown in Figure 5(c).

From the present experimental study, we infer that doping hematite nanoparticles with a large S/V ratio and higher Co concentration (greater than the threshold of 2%) might be crucial in tuning the sensitivity of the probe. In this study, however, no significant change in sensitivity was observed for the Co-doped sample with 2% doping. On the other hand, the lower LoD observed in the Co-doped hematite sample suggests enhanced interaction strength between the Co-doped hematite surface and glucose molecules. Nevertheless, irrespective of Co doping concentration, morphological, or optical properties, it remains essential to explore the interaction between glucose molecules and the hematite surface (both pristine and Co-doped) at a more fundamental level. Such insights are critical for advancing glucose detection applications. In this context, DFT calculations provide valuable theoretical insights into the inter-actions between glucose molecules and hematite surfaces, offering a deeper understanding of the material's potential for glucose sensing applications.

## 4. Theoretical analysis using DFT

### 4.1. Computational Methodology

The geometrical optimizations and electronic structure calculations of glucose molecule adsorbed on pristine and Co-doped hematite surfaces are carried out using first principles density functional theory as implemented in Vienna ab initio Simulation Package (VASP) [56, 57, 58]. For the exchange-correlation functional, Perdew-Burke-Ernzerhof (PBE), a version of generalized gradient approximation (GGA) was taken into account within projector augmented wave (PAW) method [59, 60]. The kinetic energy cutoffof 500 eV was considered for the expansion of plane wave basis sets. The Brillouin zone sampling was made in Gamma-centered method using the 3×3×1 and 7×7×1 k-point grids for self-consistent and density of states calculations, respectively, for the alpha-hematite surface. We included the Hubbard type on-site Coulomb potential U with the generalized gradient approximations (GGA+U) in the calculations due to the presence of Fe and Co atoms. Using Dudarev and Botton approach [61], we have considered U=4.0 eV due to the presence of d-orbitals in Fe and Co atoms. Additionally, using Bader charge analysis [62], the distribution and transfer of charges between the atoms in various systems were determined. All the structures were optimized until the force on each atom became less than 0.01 eV/Å, and the threshold for energy convergence between two successive steps was set at $1\times10^{-4}$ eV. In the present work, slab model of (001) surface of hematite was designed to study the adsorption of glucose molecules. The supercell consisting of hematite slab and glucose molecule was measured to be 10.07Å×10.07 Å in the XY direction which is equal to the size of 4-unit cells of $Fe_2O_3$(001). This size can give adequate space to prevent interactions between the images of glucose molecules created due to the periodicity. Along Z-direction, a vacuum of 15 Å was chosen to avoid interactions between vertical images (Figure 6a). Furthermore, we have doped two Co atoms in place of two Fe atoms in the hematite slab in order to study the effect of Co doping on glucose adsorption (Figure 6b).

### 4.2. Adsorption and Charge Transfer Analysis

The optimized structures of glucose molecule adsorbed on pristine and Co-doped hematite (001) surface are shown in Figure 6. The adsorption energy ($E_{ads}$) of pristine or doped hematite (001) surface to the glucose molecules can be calculated as follows:

$$E_{ads} = E_{surface+glucose} - E_{surface} - E_{glucose},$$

where, $E_{surface+glucose}$ is the total energy of the pristine/doped hematite surface after glucose molecule adsorption. $E_{surface}$ and $E_{glucose}$ are the energies of the pristine/doped hematite surface and glucose molecule, respectively. More the negative value of adsorption energy, more easily the molecule can be adsorbed on the surface. From the calculation, we have found that the adsorption energy of glucose molecule on pristine hematite surface is –0.24 eV whereas the value is –1.28 eV for Co-doped hematite surface. Compared to a pristine hematite surface, these values imply that the Co-doped hematite surface has the potential to be an effective glucose molecule sensor.

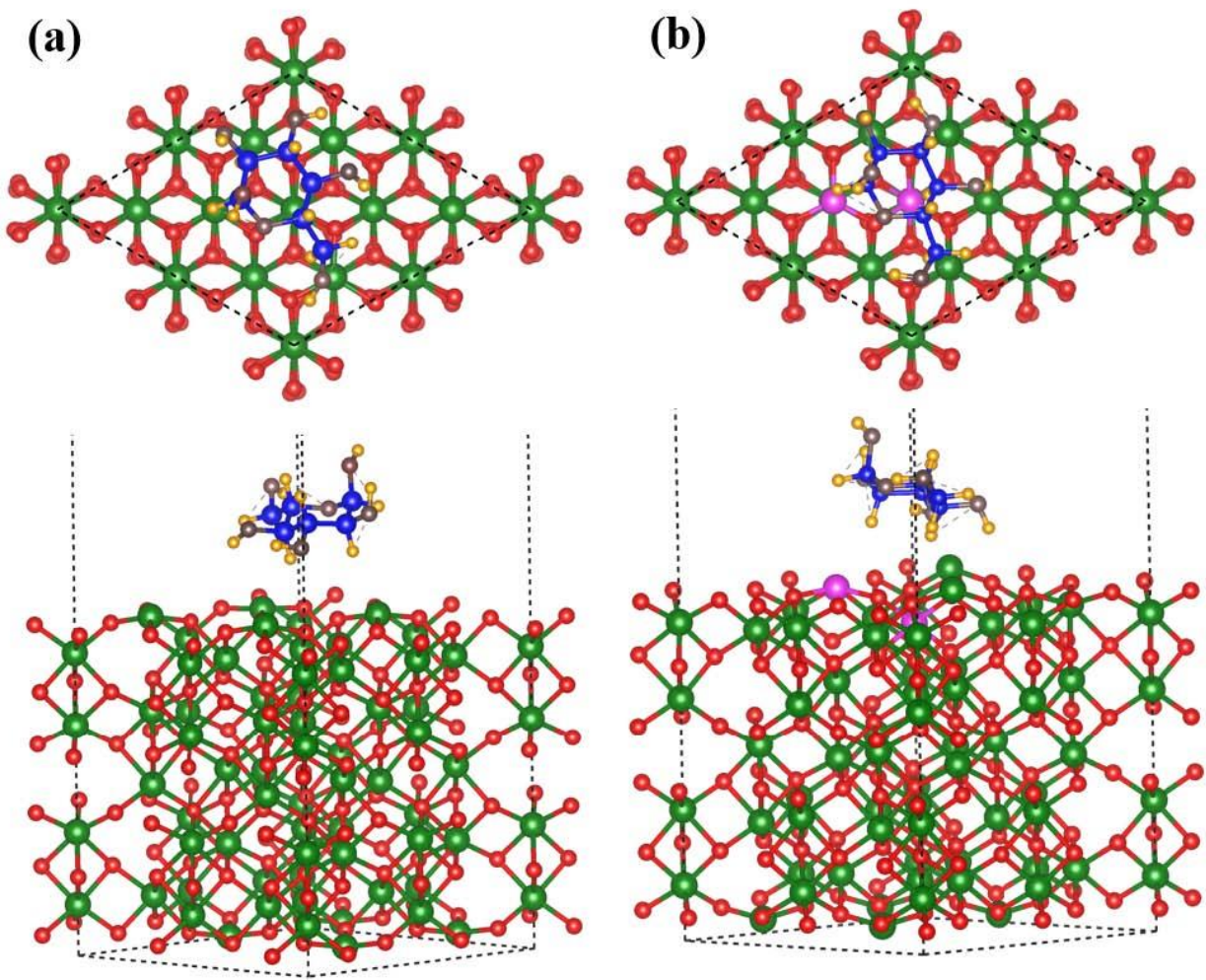

*Figure 6: Top and side views of optimized structures of glucose molecule adsorbed on (a) pristine (b) Co-doped hematite surface. (Fe, O (hematite), Co, C, O (glucose), and H atoms are represented by green, red, magenta, blue, brown and orange spheres respectively).*

Next, we have computed the charge density difference [$\Delta\rho(r)=\rho_{surface+glucose} - \rho_{surface} - \rho_{glucose}$] to visualize the electron transfer between the surface and the adsorbate in order to explain the differences in the interaction strengths of the glucose molecule in two different surfaces. The co-doped hematite surface exhibits a notable charge depletion region in comparison to the pristine hematite surface, suggesting a substantial degree of charge redistribution (Figure 7). Electrons are mostly accumulated around C and O atoms of the glucose molecule which are close to the surface. Stronger interaction between the adsorbate molecule and the surface is indicated by higher accumulation density as observed in case of doped surface. The primary sites of electron depletions are the H atoms that are connected to the C or O atoms and located close to the surface. We have used Bader charge calculations to examine the amount of charge transfer between the glucose molecule and the pristine (or Co-doped) hematite surface. The average charge of the O atoms in the glucose molecule is -1.13e, while the average charge of the O atoms on the surface is -1.17e in both systems. However, the charge of Fe atoms ranges from 1.6e to 1.8e. Charge redistribution in the Co-doped surface is caused by the average loss of 1.37e of charge experienced by Co atoms. Consequently, the entire glucose molecule loses essentially little charge when adsorbed above the pristine hematite surface, while 0.1e of a charge is transferred from the entirely glucose molecule to the Co-doped hematite surface.

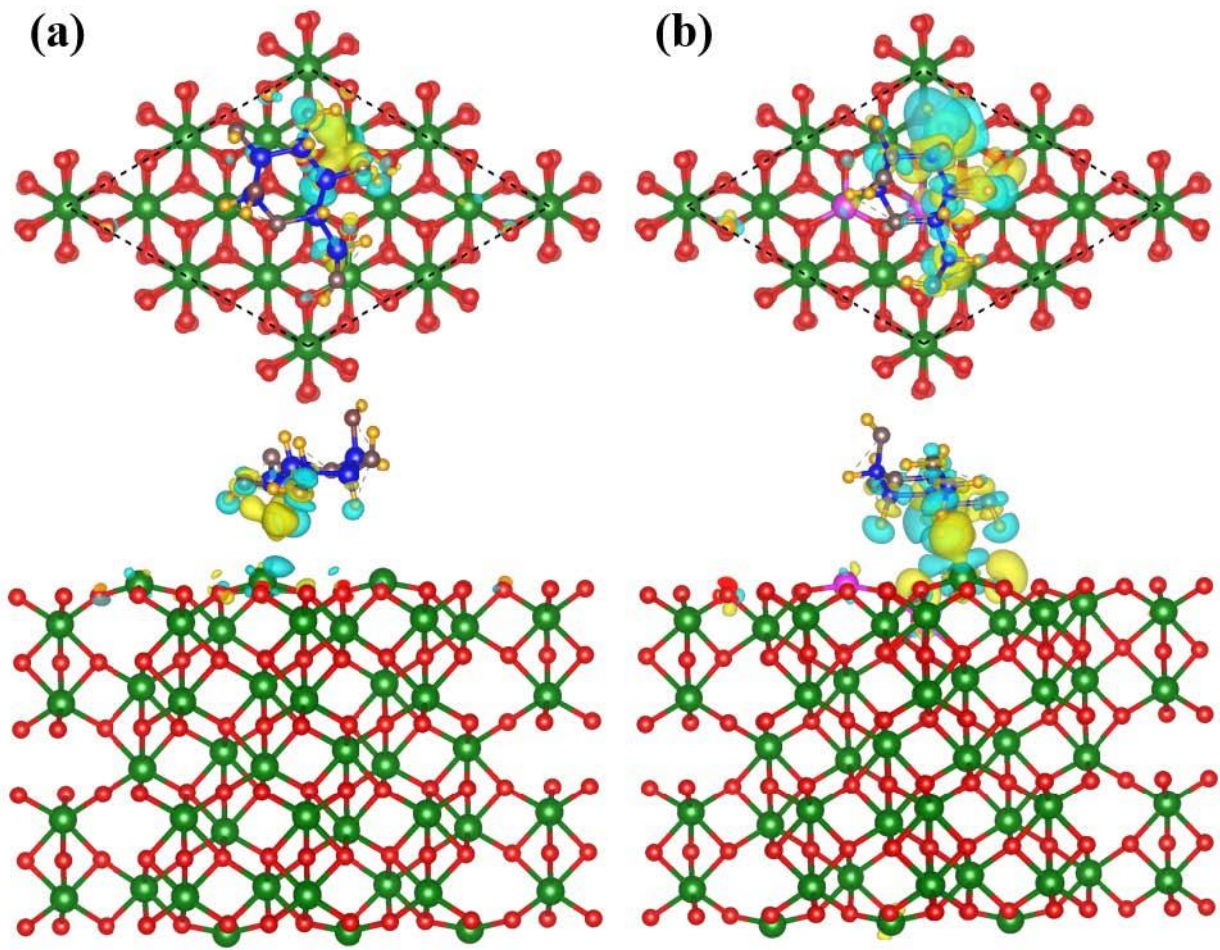

*Figure 7: Top and side views of charge density differences of glucose molecule adsorbed on (a) pristine (b) Co-doped hematite surface. Cyan and yellow colors represent the electron depletion and accumulation regions respectively. The values of iso-surface are taken 0.0005 e/Å³ and 0.001e/Å³ for (a) and (b) respectively for better visualization.*

To get more insight into the interactions between orbitals of the adsorbate and the surface, we have calculated spin-polarized total and projected density of states (PDOS) of adsorbed glucose molecule on pristine and Co-doped hematite surfaces as depicted in Figure 8. In both systems, the d-states of Fe atoms predominate in the deeper valence band regions, while the O-2p states of the hematite surface contribute primarily to the total density of states close below the Fermi level. When a glucose atom is adsorbed on the surface, 1s states of H atoms and 2p states of C atoms from the glucose molecule are get hybridized with Fe-3d and O-2p states from the hematite surface in the energy level below 6 eV from Fermi level for spin-up channel. But for spin-down, the only contribution to the DOS is only from glucose molecule in this energy range. But, for the spin-down channel, orbitals of atoms from pristine surfaces occupy the majority of the conduction band regions. On the other hand, Figure 8b shows a slight contribution from the Co-3d state in spin-up DOS for the Co-doped hematite surface, which could improve the interaction with the glucose molecule. Furthermore, the majority of the DOS in the energy range of –1 eV to –6 eV was created by the hybridization of the glucose molecule's O-2p, C-2p, and H-1s states with the (O-2p, Fe-3d, and Co-3d states) of undoped (Co-doped) hematite surfaces, which led to the molecule's binding to the surfaces. The weaker interaction is indicated by the more localized states of the glucose molecule within this energy range for the pristine hematite surface. Conversely, the O-2p, C-2p states of glucose molecule are getting delocalized when the hematite surface is doped with Co atoms. The glucose molecule interacts more and binds to the doped surface firmly as a result of these orbitals spreading [63, 64]. This observation correlates with the improved LoD observed in the $F_{Co\text{-}doped\text{-}hematite}$ probe. So, overall, our combined experimental and theoretical work outlines the potential of Co-doped hematite for further exploration in glucose sensor technologies.

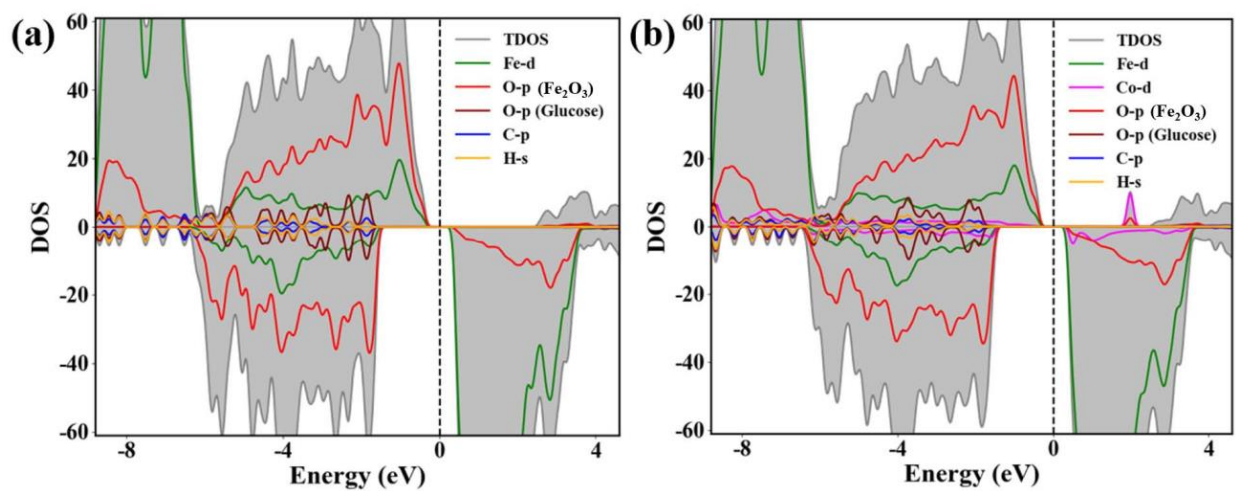

*Figure 8: Total and projected density of states of glucose molecule adsorbed above (a) pristine (b) Co-doped hematite surface. Fermi level is set at 0 eV.*

## 5. Conclusions

We investigated the glucose sensing capabilities of pristine and Co-doped hematite synthesized via the hydrothermal method. Structural, morphological, and optical characterizations were thoroughly conducted using X-ray diffraction (XRD), scanning electron microscopy (SEM), and UV-Visible spectroscopy. The fiber-optic evanescent wave (FOEW) setup was used to evaluate the glucose sensing performance, revealing that while the sensitivity remained similar for both the pristine and Co-doped hematite samples, the Limit of Detection (LoD) showed a significant improvement in the Co-doped sample. Specifically, the LoD values were found to be 6.12 mM for the pristine sample and 3.99 mM for the Co-doped hematite sample, indicating that Co doping enhances the material's interaction with glucose molecules. To support these experimental findings, Density Functional Theory (DFT) calculations were employed to explore the adsorption mechanisms and charge transfer processes at the molecular level. The DFT analysis revealed a substantial increase in adsorption energy, with values of -0.24 eV for the pristine surface and -1.28 eV for the Co-doped surface, indicating stronger glucose binding on the Co-doped hematite. Additionally, charge density difference analysis and projected density of states (PDOS) showed enhanced charge transfer and orbital delocalization upon doping, which correlates with the improved LoD observed experimentally. Overall, this study demonstrates that while Co doping did not significantly impact sensitivity within the FOEW setup, it played a critical role in lowering the LoD, thus improving the glucose detection capabilities. Future research could explore higher Co doping concentrations or alternative sensor configurations to further enhance sensitivity. This work contributes to the ongoing investigation of hematite-based materials in biosensing technologies, illustrating the value of integrating experimental and theoretical approaches to advance glucose sensor development.